# Symmetric Two Dimensional Photonic Crystal Coupled Waveguide with Point Defect for Optical Switch Application

H. Hardhienata, H. Alatas
*Theoretical Physics Division, Departement of Physics Bogor Agricultural University*
JL. Meranti, Ged.Wing S, Kampus IPB Darmaga
Bogor , Indonesia

Email: hendradi@ipb.ac.id

**Abstract-** Two dimensional (2D) PC's are well known for its ability to manipulate the propagation of electromagnetic wave inside the crystal. 1D and 2D photonic crystals are relatively easier to fabricate than 3D because the former work in the microwave and far infrared regions whereas the later work in the visible region and requires smaller lattice constants. In this paper, simulation for a modified 2D PC with two symmetric waveguide channels where a defect is located inside one of the channel is performed. The simulation results show that optical switching is possible by modifying the refractive index of the defect. If more than one structure is applied this feature can potentially be applied to produce a cascade optical switch.

## I. Introduction

The ability to control the electric properties of material has lead to many theoretical and practical advances that we still use today. This included the development of semiconductor physics which then triggered the transistor revolution that has changed the technological trends in our world dramatically as is evident by our dependences in electronic products. However, in the last few decade intense effort to control the optical properties in a material through deeper study on the interaction between radiation field and matter brought birth to a new and very promising scientific frontier. The ability to control and localize the propagation of light in a material has given human new technological benefit, to mention a few: fiber optics, laser, and high speed communication have already changed the future trends in information technology.

In 1987 E. Yablonovich and S. John independently performed intensive and systematic study on the characteristics of an electromagnetic wave propagating inside a photonic crystal. Their research has opened many important scientific discoveries and application in information technology. A photonic crystal (PC) is a material with dielectric periodicity and is able to control light propagation. A photonic crystal with two different refractive indices can be formed to produce 1D, 2D, and 3D photonic crystals. 1D and 2D photonic crystals are relatively easier to fabricate than 3D because the former work in the microwave and far infrared regions whereas the later work in the visible region and requires smaller lattice constants. The scale of the lattice constants is however usually not in the order of angstroms as is the case for ordinary crystals but are more likely in the order of the electromagnetic wavelength that is propagating inside the crystal such as 1 µm or less for visible light and 1 cm for microwaves [1].

The analogy in optical engineering of a semiconductor as a guiding material for electrons is the photonic crystal. Here, the periodic structure of dielectric material that comprises a photonic crystal acts as a periodic potential rather than the atoms in a semiconductor. Therefore it should not be too surprising if similar phenomenon in semiconductor also occurs for photonic crystals. In fact band gaps due to Bragg diffraction does also occur in photonic crystals so that certain frequencies passing through the crystal will not be transmitted [2]. If for some frequency range and any incident angle, all the lights are reflected by a photonic crystal then it has the property of a complete photonic band gap [3]. The existence of defect layers between periodic structures in a 1D photonic crystal will produce a permitted transmittance frequency (pass band) inside the band gap whose characteristics can be altered by modifying the defect properties (e.g. index or length of the defect). Furthermore, we previously reported in [4] that inclusions of two defect layers inside a periodic 1D photonic crystals enables frequency shifting and peak transmittance control of the pass band by modification of the refractive index of the two defects.

## II. Basic Theory

The derivation of electromagnetic (EM) wave propagating inside a photonic crystal can be obtained directly from Maxwell's equation with the help of the constitutive equations which describes the response of an EM wave inside a material.

Using vector calculus, the vectoral wave equation for the electric field $\vec{E}$ then takes the form of

$$\vec{\nabla} \times \vec{\nabla} \times \vec{E}(\vec{r}) - \omega^2 \mu \varepsilon \vec{E}(\vec{r}) = 0 \quad (1)$$

for a homogenous, isotropic, and source free medium. Using the relation $k = \omega\sqrt{\mu\varepsilon}$ equation (1) becomes:

$$\vec{\nabla} \times \vec{\nabla} \times \vec{E}(\vec{r}) - k^2 \vec{E}(\vec{r}) = 0 \quad (2)$$

Now, consider scattering case where the scatter has a dielectric function $\varepsilon(r)$ and is located at an infinite background medium $\varepsilon_B$. When the background medium is not a vacuum and the material is nonmagnetic, the total electric field from an incident wave c that is propagating in the background medium (including the scattering plus incident field) is then given by the solution of the vectorial wave equation

$$\vec{\nabla} \times \vec{\nabla} \times \vec{E}(\vec{r}) - k_0^2 \varepsilon(r) \vec{E}(\vec{r}) = 0 \quad (3)$$

where $k_0 = k\sqrt{\varepsilon(\vec{r})} = \omega^2/c^2$ is the vacuum wave number. Using the dielectric contrast to ease the next steps

$$\Delta\varepsilon(\vec{r}) = \varepsilon(\vec{r}) - \varepsilon_B \quad (4)$$

the homogenous vectorial wave now can be written in inhomogenous form

$$\vec{\nabla} \times \vec{\nabla} \times \vec{E}(\vec{r}) - k_0^2 \varepsilon_B \vec{E}(\vec{r}) = k_0^2 \Delta\varepsilon(\vec{r}) \vec{E}(\vec{r}) \quad (5)$$

### III. GREEN'S TENSOR FOR 2D PC

Here we consider the case of a transverse magnetic (TM) EM mode where the magnetic field is perpendicular to the plane of incidence inside a 2D photonic crystal square rod. A square rod consists of regular array of dielectric slabs as depicted in Figure 1.

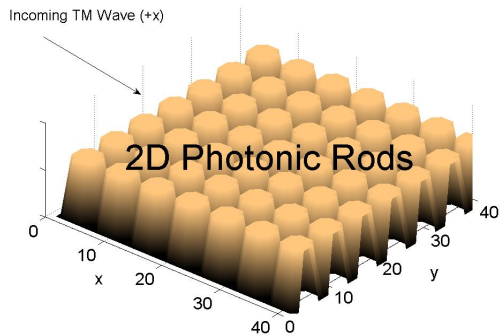

Fig. 1. 2D PC square rod scattered by an incoming TM wave from the left.

Green's function of a wave equation is a solution of a point source wave equation. If the solution of a wave equation due to a point source is known, then the solution of several point sources can be found by linear superposition of each source point. To obtain the scalar wave equation solution for one source point one can follow the derivation in [5]. For the wave equation

$$(\vec{\nabla}^2 + k^2)\vec{E}(\vec{r}) = s(\vec{r}) \quad (6)$$

one can find a Green's function that is a solution of

$$(\vec{\nabla}^2 + k^2)g(\vec{r},\vec{r}') = -\delta(\vec{r} - \vec{r}') \quad (7)$$

Calculation using equation (6) and (7), the Green's scalar function for a source point can be obtained in the form of

$$g(\vec{r}) = \frac{1}{4\pi|\vec{r}-\vec{r}'|} e^{ik|\vec{r}-\vec{r}'|} \quad (8)$$

This is the spatial representation of the dyadic Green's function. Furthermore, the electric field $\vec{E}(r)$ is related to the current density $\vec{J}(r)$ by the equation

$$\vec{E}(\vec{r}) = i\omega\mu \int_V \vec{\vec{G}}(\vec{r},\vec{r}') \cdot \vec{J}(\vec{r}') d\vec{r}' \quad (9)$$

which is a solution of

$$\vec{\nabla} \times \vec{\nabla} \times \vec{E}(\vec{r}) - k_0^2 \vec{E}(\vec{r}) = i\omega\mu \vec{J}(\vec{r}) \quad (10)$$

substituion of equation (3.1.5) to (3.1.4) gives

$$\vec{\nabla} \times \vec{\nabla} \times \vec{\vec{G}}(\vec{r},\vec{r}') - k_0^2 \vec{\vec{G}}(\vec{r},\vec{r}') = \bar{I}\delta(\vec{r}-\vec{r}') \quad (11)$$

where $\bar{I}$ is called the unit dyad. Using equation (9), (10) and (11) and the fact that the field is related to the Green's tensor by [5]

$$\vec{E}(\vec{r}) = i\omega\mu \int_V d\vec{r}' \vec{\vec{G}}(\vec{r},\vec{r}') \vec{J}(\vec{r}') \quad (12)$$

then $\overline{\overline{G}}(r,r')$ can be derived to have the form of

$$\vec{\vec{G}}(\vec{r},\vec{r}') = \left[\bar{I} + \frac{\vec{\nabla}\vec{\nabla}}{i\omega\varepsilon}\right] g(\vec{r},\vec{r}') \quad (13)$$

which is called the 3D Green Tensor

## IV. 2D GREEN'S FUNCTION

The derivation of the scalar 2D Green's function can be performed by starting from the physical point of view of this tensor. The 2D Green's tensor is the field generated in an observation plane where (for this case) z = constant by an infinite line source that is extending in the $z$ direction in addition of having a $\exp(ik_z z)$ dependence. This means the Green's function can be obtained by summing all point source that are placed in line or in other words integration over this line source:

$$g_{2D}^B(\vec{r},\vec{r}\,') = \int_{-\infty}^{\infty} dz' \, g_{3D}^B(\vec{r},\vec{r}\,') e^{ik_z z} \quad (14)$$

with the relative coordinate $\rho^2 = x^2 + y^2$ and assumption that the line source is located at $x' = y' = 0$ then

$$g_{2D}^B(\vec{r},\vec{r}\,') = \int_{-\infty}^{\infty} dz' \frac{\exp[ik_B\sqrt{\rho^2 + (z-z')^2}]}{4\pi\sqrt{\rho^2 + (z-z')^2}} e^{ik_z z}$$

$$= \frac{i}{4} H_0(k_a, a) \exp[ik_z z] \quad (15)$$

Where $H_0$ is known as Hankel Function of the first kind.

In the considered case, namely the TM polarization, Green's tensor $G^B(\vec{\rho},\vec{\rho}\,')$ reduces to a scalar Green's function with z = 0, hence

$$G^B(\vec{\rho},\vec{\rho}\,') = G_{zz}^B \quad (16)$$

Because here the incoming wave is propagating in the x – y plane then $k_z = k_B$ so that [7]:

$$\vec{\vec{G}}^B{}_{zz} = \frac{i}{4} H_0(k_\rho, \rho) \quad (17)$$

## V. NUMERICAL CALCULATION

The Greens's scalar function for a point source can be obtained from [5] namely

$$g(r) = \frac{1}{4\pi|\mathrm{r\text{-}r'}|} e^{ik|\mathrm{r\text{-}r'}|} \quad (18)$$

Clearly $\vec{\vec{L}}$ is zero for the TM case $(\vec{\vec{L}}_{zz} = 0)$. The Lippman-Schwinger equation can be solved numerically by using its discretized version (other approach can be made by finite elements). In doing so a grid that consists of N meshes are constructed. Each mesh has a constant volume of $V_i$.

$$\vec{E}(\vec{r}) = \vec{E}^0(\vec{r}) + \lim_{\delta V \to 0} \int_{V-\delta V} d\vec{r}\,' G^B(\vec{r},\vec{r}\,') \cdot k_0^2 \Delta\varepsilon(\vec{r}\,')\vec{E}(\vec{r}\,') - \vec{\vec{L}} \cdot \frac{\Delta\varepsilon(\vec{r})}{\varepsilon_B} \vec{E}(\vec{r}) \quad (20)$$

$$\vec{E}_i = \vec{E}_i^0 + \sum_{j=1, j \neq 1}^{N} \vec{\vec{G}}_{ij}^B \cdot k_0^2 \Delta\varepsilon_j \vec{E}_j V_j + \vec{\vec{M}} \cdot k_0^2 \Delta\varepsilon_i \vec{E}_i - \vec{\vec{L}} \cdot \frac{\Delta\varepsilon_i}{\varepsilon_B} \vec{E}_i \quad (21)$$

Note that the equation is singular when the observation source is located in the source point $(\vec{r} = \vec{r}\,')$. Therefore the Green's tensor is also singular for this case.

To overcome this difficulty whenever $\vec{r} = \vec{r}\,'$ are inside the scatter volume the singularity of Green's tensor has to be treated separately. This singularity problem can be solved using the Lippman-Schwinger equation (20). Here the integral excludes the singularity with the consequence of adding a source dyadic term $\vec{\vec{L}}$ which depends on the shape of the exclusion [6]:

$$\vec{\vec{L}}_i = \begin{pmatrix} \frac{1}{2} & 0 & 0 \\ 0 & \frac{1}{2} & 0 \\ 0 & 0 & 0 \end{pmatrix} \quad (19)$$

The discretized version is given in equation (21) where $\vec{E}_i = \vec{E}(r_i)$ is the discretized field and $\varepsilon_i = \varepsilon(\vec{r}_i)$ the discretized dielectric contrast. The discretion need not be regular but can vary locally to enhance the required accuracy, with a smaller mesh where the dielectric contrast $\Delta\varepsilon(\vec{r})$ is large [7]. Here the so called self-term $\vec{\vec{M}}$ occurs due to the discretization from an integration into summation. The self term is less significant then the other terms and therefore it is neglected by several researchers. However it plays an important role in the accuracy of the computation.

Equation (20) is non singular when the observation point $\vec{r}$ is located outside the scatter because the integral is only limited to the scatter volume. Furthermore, the field at any point in the background is entirely determined from the field inside the scatter. Therefore the field inside the scatter can be determined first and the field outside the scatter calculated afterwards by using the information of the inside field. This procedure can decrease calculation error and is applied in the program building [8].

The value of the self-term $M_i$ can be computed by [7]:

$$\vec{\vec{M}}_i = \lim_{\delta V \to 0} \int_{V_i - \delta V} d\vec{r}' \vec{\vec{G}}^B(\vec{r}_i, \vec{r})  \quad (22)$$

More explicitly in terms of components

$$\vec{\vec{M}}_i = \begin{pmatrix} \frac{i\pi}{4}\alpha\gamma & 0 & 0 \\ 0 & \frac{i\pi}{4}\alpha\gamma & 0 \\ 0 & 0 & \frac{i\pi}{4}\beta\gamma \end{pmatrix} \quad (23)$$

where

$$\alpha = \left(2 - \frac{k_\rho^2}{k_B^2}\right); \beta = \left(1 - \frac{k_z^2}{k_B^2}\right); \quad (24)$$

and

$$\gamma = \left(\frac{R_i^{eff}}{k_\rho} H_1(k_\rho, R_i^{eff}) + \frac{2i}{\pi k_\rho^2}\right) \quad (25)$$

## V. RESULTS AND DISCUSSIONS

Using the numerical equation (19) the electric field distribution for an incoming TM wave on a dielectric scatter can be simulated. Previous simulation using this method for TM scattering on a 2D single rod holds very similar result with the analytical results. Here we concentrate on TM EM wave scattering from the right x axis on a 2D PC symmetric waveguide with a single point defect located on the second waveguide as depicted in Fig. 2.

For computational reason the 2D PC is limited to 7 x 7 square rods with similar lattice distance. To allow EM wave traveling inside the structure, the scattered index is set to be 3.67 and is surrounded by air (n = 1). The wavelength of the incoming TM wave is $2.85 \cdot 10^{14}$ which is a traveling wave mode for this 2D configuration [1]. The radius of the scattered rods is 0.25 μm.

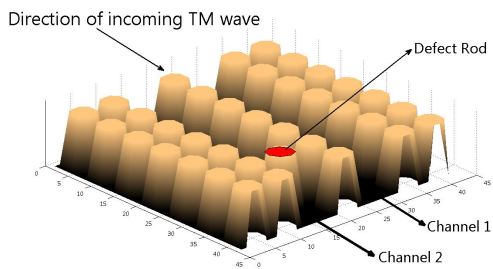

Fig. 2. Configuration of the considered system

Next we calculate the electric field distribution inside the 2D photonic crystal and calculate the intensity at x = 0 and x = L for all y lattice points. We call the former as the input intensity and the later the output intensity. The simulation is performed by varying the refractive index of the point defect for example by using a liquid crystal whose index can be altered, and then study the intensity profile at the two ends. First we consider the case where there is no defect at all. As expected, the electric waveguide profile shows a symmetrical and even distribution inside each channel as depicted in Fig. 3.

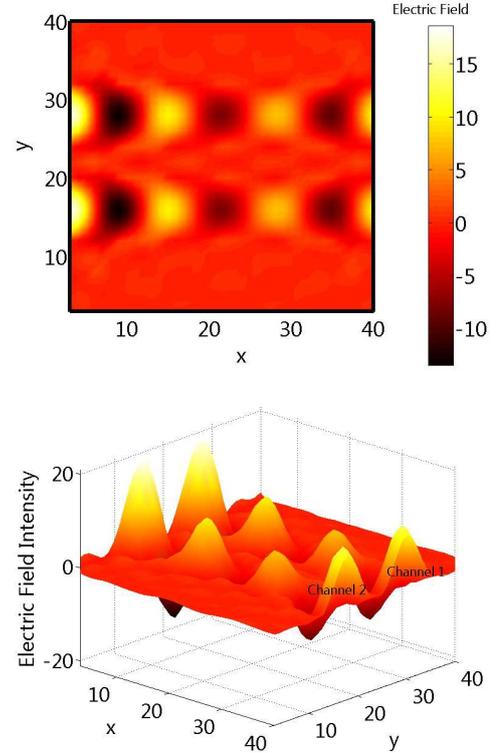

Fig. 3. Electric Field Profile for considered model with no Defects inside both Channels in (a) 2D representation and (b) 3D representation. Note the symmetrical and similar distribution of the electric field profile inside the channels/

It can be seen in Fig. 4 that there is a loss in intensity of about 60% between the two edges without any phase shift. The bandwidth of the incoming and outgoing intensity of both channels is conserved.

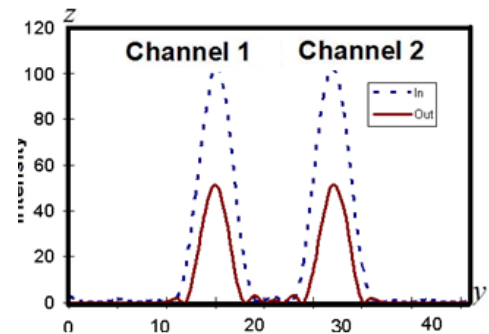

Fig. 4. Intensity profile for considered model with no Defects inside the Channel. The left depicts the intensity at channel 1 whereas the right depicts channel 2. The blue dashed line shows the input (x = 0) intensity whereas the red solid line the output intensity (x = L)

The decrease in transmittance intensity as it travels inside the channels is due to the amount of rods used in the simulation (7 x 7) so that parts of the travelling wave leak out to the upper (y axis) and lower (-y axis) ends of the structure. This loss can be minimized if the simulation is performed for a structure with higher amount of rod arrangement. However more rods will require more memory, hence longer calculation time.

Second, we set the point rod defect index to be 1.41 and repeat the simulation. Fig. 5 shows that the defect causes a phase shift of the intensity between the first and second channel. This phase shift resulted in the vanishing of intensity in channel 2 whereas the intensity in the first channel is only shifted lightly and decreases little from the input.

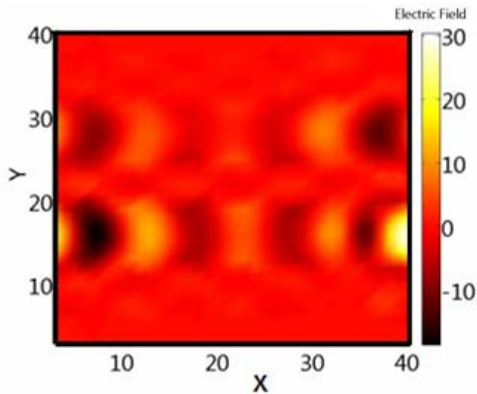

Fig. 5. Electric Field Profile for a point Defect 1.41 Inside Channel 2. Note that there is a phase shift between the wave travelling at channel 1 (above) and channel 2 (below)

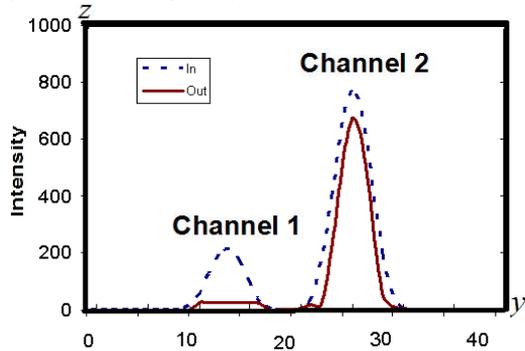

Fig. 6. Intensity profile for considered model for Defect index 1.41. The intensity in channel 1 dissaperas at the output point due to phase shift. In the other channel the intensity is only decreasing slightly

The input intensity calculated at the edges of the x-axis mesh (mesh 40) is presented in Fig. 6. The profile also differs with the previous result (Fig. 4) because the input transmittance in channel 2 is higher than in channel 1. Further theoretical study is necessary to explain this phenomenon. Nonetheless it seems that the defect reflects part of the travelling wave that interference constructively at x = 0.

Furthermore, although the EM wave seems to travel mostly through channel 2, however a look at the electric field propagation profile shows that the wave still travels through both channels. It should be pointed out once again that the disappearance of the output intensity in channel 1 is caused by the phase shift due to insertion of a defect of 1.41 rather than blocking from the defect. In the latter explanation it will be described that for higher defect index the wave blocking or scattering will become more apparent. Another result that has to be stated is that the intensity magnitude in this case is higher than for the case where there is no defect. Further theoretical study has to be performed to fully understand this phenomenon however it seems to be clear that the defect produces this result. A reasonable explanation is that the defect absorbs part of the travelling wave in channel 1 by coupling effect so that the intensity in channel 2 increases significantly.

An interesting phenomenon occurs when we set the point rod defect index to 1.73. Fig. 7 shows that instead vanishing out the waveguide intensity in channel 2 as was in the previous case the defect produces coupling phenomenon, namely absorbing the traveling wave field in channel 1. Therefore, the output intensity in channel 2 now becomes very high compared to the intensity in the other channel. Once again the magnitude of the intensity is higher than if there were no defects but still lower than if the defect rod is set for 1.41.

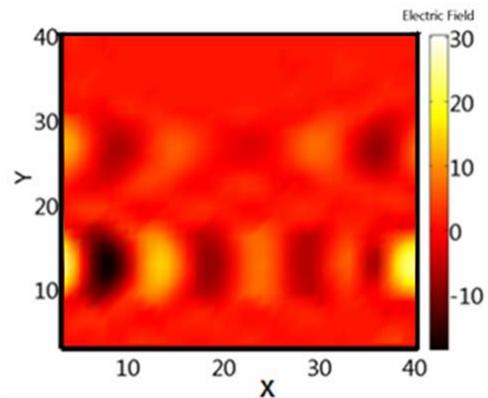

Fig. 7. Electric Field Profile for a point Defect 1.73 Inside Channel 2. Note the disappearance of the field in the first channel.

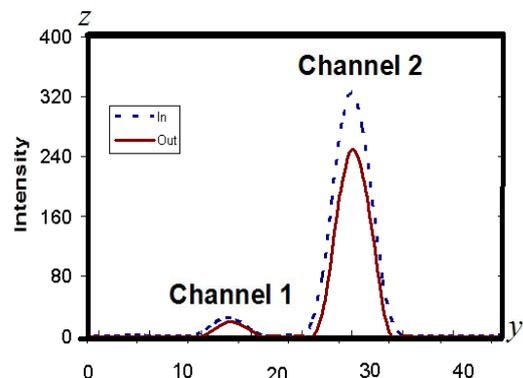

Fig. 8. Intensity profile for considered model for Defect index 1.73

Finally, the point defect rod index is further increased to 2.64. This time the defect reflex almost all the wave that travelling through channel 2 as given in Fig. 9. Furthermore the input wave in channel 2 is higher than in channel 1 as depicted in Fig. 10.

This feature is similar to the result for the previous defect index set to 1.41 or 1.73 but this time channel 1 carries most of the field. It should be emphasized that the loss of the output intensity in channel 2 is caused by defect blocking rather than phase shifting. This is evident by the gradual decrease of the electric field in channel 2.

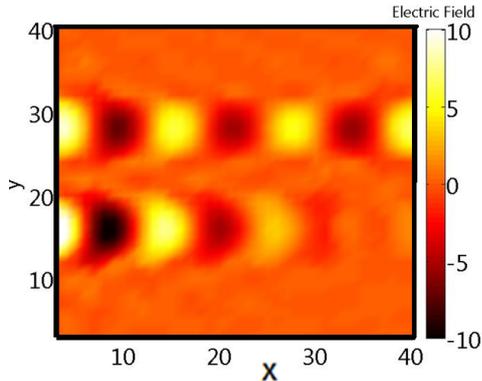

Fig. 9. Intensity profile for considered model for Defect index 2.64. Note the gradual decrease of the field in channel 2.64.

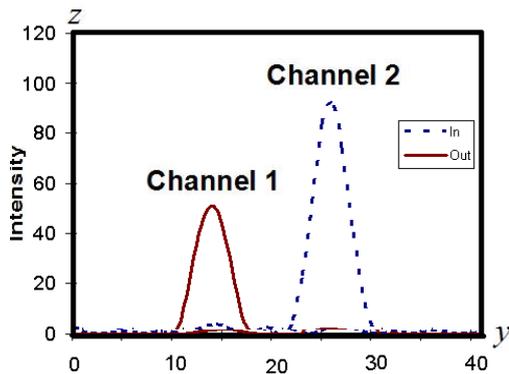

Fig. 10. Intensity profile for considered model for Defect index 2.64

## VII INSTRUMENTATION SCHEME

PC rods or slabs are not only easier to fabricate but are also able in realizing various control of light waves. It is believed that such systems are suitable for miniaturization of individual optical devices thus making densely integrated photonic circuits possible [9]. The unique coupling-decoupling feature can be applied to produce a defect index based cascade optical switcher by combining the output of one structure into the input of the other structure. The instrument arrangement with 4 outputs is depicted in Fig. 11. Three 2D photonic crystal optical switches added with a changeable single defect rod can produce 4 channels where the user can choose where the wave should travel through by proper setting of the defect indexes.

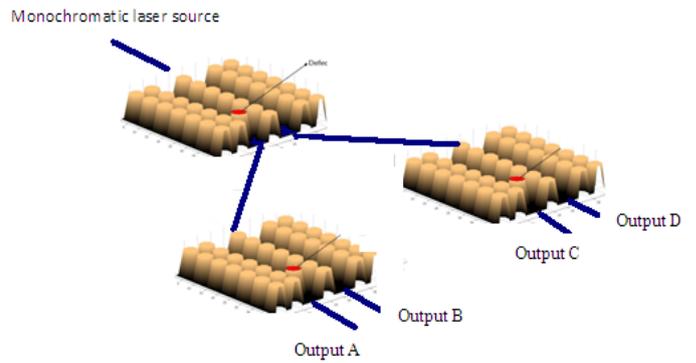

Fig. 11. Instrumentation Scheme for the Multiple Optical Switching

## VII CONCLUSION

From the results it can be concluded that optical switching between the two channels is possible by altering the index of the point defect. Transmittance on both channels can be switched to channel 2 only, by setting the defect to 1.73 which produces field couplings. In the other hand, transmittance can be altered to channel 1 only, this time by giving a high index for the defect, in this case 2.64. Alternation of the defect index can be performed by using liquid crystals as defect material.